\documentclass[journal,transmag]{IEEEtran}

\usepackage{lineno}
\usepackage{amsmath,graphicx,makecell}
\usepackage{changes}
\usepackage{enumitem}
\graphicspath{{./figs/}}
\usepackage{threeparttable}
\usepackage{booktabs}
\usepackage{algorithmic}
\usepackage[linesnumbered, ruled, vlined]{algorithm2e}
\usepackage{caption}
\usepackage{subcaption}
\usepackage{url}

\begin{document}
\title{Impact of Climate Simulation Resolutions on Future Energy System Reliability Assessment: A Texas Case Study}

\author{\IEEEauthorblockN{Xiangtian Zheng,\IEEEauthorrefmark{1} 
Le Xie,\IEEEauthorrefmark{1,3,$\dagger$}
Kiyeob Lee,\IEEEauthorrefmark{1} 
Dan Fu,\IEEEauthorrefmark{2}
Jiahan Wu,\IEEEauthorrefmark{1}
and Ping Chang\IEEEauthorrefmark{2}}
\IEEEauthorblockA{\IEEEauthorrefmark{1}Department of Electrical and Computer Engineering, Texas A\&M University, College Station, TX 77843 USA}
\IEEEauthorblockA{\IEEEauthorrefmark{2}Department of Oceangraphy, Texas A\&M University, College Station, TX 77843 USA}
\IEEEauthorblockA{\IEEEauthorrefmark{3}Texas A\&M Energy Institute, College Station, TX 77843 USA}
\IEEEauthorblockA{\IEEEauthorrefmark{$\dagger$}Corresponding author}
}

\markboth{iEnergy, 2023}%
{Zheng \MakeLowercase{\textit{et al.}}: Impact of Climate Modeling Resolution on Power System Reliability Assessment}

\IEEEtitleabstractindextext{%
\begin{abstract}
The reliability  of energy systems is strongly influenced by the prevailing climate conditions. With the increasing prevalence of renewable energy sources, the interdependence between energy and climate systems has become even stronger. This study examines the impact of different spatial resolutions in climate modeling on energy grid reliability assessment, with the Texas interconnection between 2033 and 2043 serving as a pilot case study. Our preliminary findings indicate that while low-resolution climate simulations can provide a rough estimate of system reliability, high-resolution simulations can provide more informative assessment of low-adequacy extreme events. Furthermore, both high and low-resolution assessments suggest the need to prepare for severe blackout events in winter due to extremely low temperatures.
\end{abstract}

}

\maketitle

\IEEEdisplaynontitleabstractindextext

\IEEEpeerreviewmaketitle

\section{Introduction}
Climate change has emerged as one of the most pressing global challenges, and the energy sector is a key player in enabling decarbonization. However, climate change also poses significant challenges to the energy sector, such as ensuring the availability of energy resources~\cite{wu2021open}, accommodating the expansion of electricity demand~\cite{auffhammer2017climate}, and enhacing the resilience of power networks~\cite{panteli2015influence}. To address these challenges and ensure a sustainable and resilient energy future, it is crucial to develop a reliable and rigorous approach to model and assess the impacts of climate change on power systems.

Recent studies in the field of climate-energy research have analyzed the impact of climate change on energy grids in multiple aspects.
Researchers have quantified the increase in future electricity demand~\cite{van2019amplification} and the expansion of wind generation~\cite{pryor2020climate}. Meanwhile~\cite{yalew2020impacts} shows clear trends of capacity and load throughout the globe but uncertainty on the regional scale.
Studies have also investigated the influences on grid reliability and resilience at different scales.
\cite{perera2020quantifying} shows the drop of power grid reliability due to both low impact variations and extreme events, 
\cite{tong2021geophysical} shows that the reliability of electricity supply with varying renewable mix and energy storage has regional geophysical constraints.
while
\cite{zeyringer2018designing} shows that the British grid with high renewable penetration grids may suffer operational inadequacy due to single-year planning.
Furthermore,~\cite{perera2023challenges} shows a relatively high vulnerability of urban areas due to the compound effects of urban density and climate change.

These studies rely heavily on reliable climate projections. However, constructing reliable statistical prediction models based solely on historical weather data is challenging due to the limited availability of reliable observational records, particularly weather extremes, such as hurricanes. Therefore, climate simulations play a critical role in accurately projecting future climate conditions and assessing their impacts. Recent advances in climate modeling have led to significant improvements, such as increased spatial resolution~\cite{chang2020unprecedented}, improved representation of physical processes~\cite{palmer2019stochastic}, and machine learning and AI based technologies~\cite{kashinath2021physics,ragone2018computation}. In particular, the recent unprecedented set of high-resolution climate simulations~\cite{chang2020unprecedented} is capable of explicitly representing some of small-scale regional processes and yields more realistic weather extreme statistics that are not adequately simulated by the standard low-resolution climate models used by the latest assessment report of the Intergovernmental Panel on Climate Change (IPCC)~\cite{portner2022climate}. As such, these high-resolution climate simulations are particularly valuable for assessing the impacts of climate change on regional energy system reliability that requires high-resolution climate information. 

While high-resolution climate simulations show promise for improving power system adequacy assessment, there are several key research gaps that must be addressed to fully unlock its potential for the energy sector. One major challenge is the need for more comprehensive and consistent methodologies to integrate climate simulation data into power system planning, from long-term capacity and transmission expansion to short-term demand forecasting and dispatch strategies. In addition, more accurate and detailed climate information is required to improve the accuracy of modeling and simulation efforts, particularly at the local and regional level, where extreme events can play a vital role. 

In this paper, leveraging an ongoing collaborative effort in climate and energy system modeling, we conduct a pilot study on the impact of different resolutions of climate simulations on energy grid reliability assessment.\footnote{The data, model and codes for all analyses in this paper are publicly available at the {GitHub repository}~\url{https://github.com/tamu-engineering-research/ClimateResolutionPowerReliability}.} We find that while low-resolution climate simulation can aid in a crude system reliability assessment, high-resolution climate simulation can reveal a higher frequency of extreme events.
Additionally, both high and low-resolution assessments suggest the need to prepare for severe blackout events in winter due to extremely low temperatures.

It should be noted that for this preliminary study, we have made the assumption that the future Texas grid's capacity and load are independent of the projected climate change. Furthermore, our power grid simulations are solely driven by projected future climate data as input rather than interactive simulations that consider the potential impacts of the energy sector on climate change.
Besides, we only use one single simulation of high (0.25 degree or 25 km) and low-resolution (1 degree or 100 km) model under Representative Concentration Pathway (RCP) 8.5 forcing for this preliminary study.\footnote{RCP 8.5 represents a ``very high baseline emission scenario'' in which greenhouse gas concentrations are projected to continue to rise throughout the 21$^\text{st}$ century, which is generally taken as the basis for worst-case climate change scenarios.} Therefore, future studies with ensembles of simulations are needed to validate the statistical robustness of the conclusions.

The rest of the paper is structured as follows. Section~\ref{sec:problem_formulation} introduces the definition of resource adequacy, which is used as a criterion for evaluating system reliability. Section~\ref{sec:results} shows the impact of different resolutions of climate data on grid reliability, the sensitivity of grid reliability to climate conditions, and discusses the potential of climate energy research and future research directions. Section~\ref{sec:conclusion} summarizes the findings of the study and draws conclusions on the impact of climate simulation resolution on grid reliability assessment.

\begin{figure*}[hbt!]
    \centering
    \includegraphics[width =0.95\textwidth]{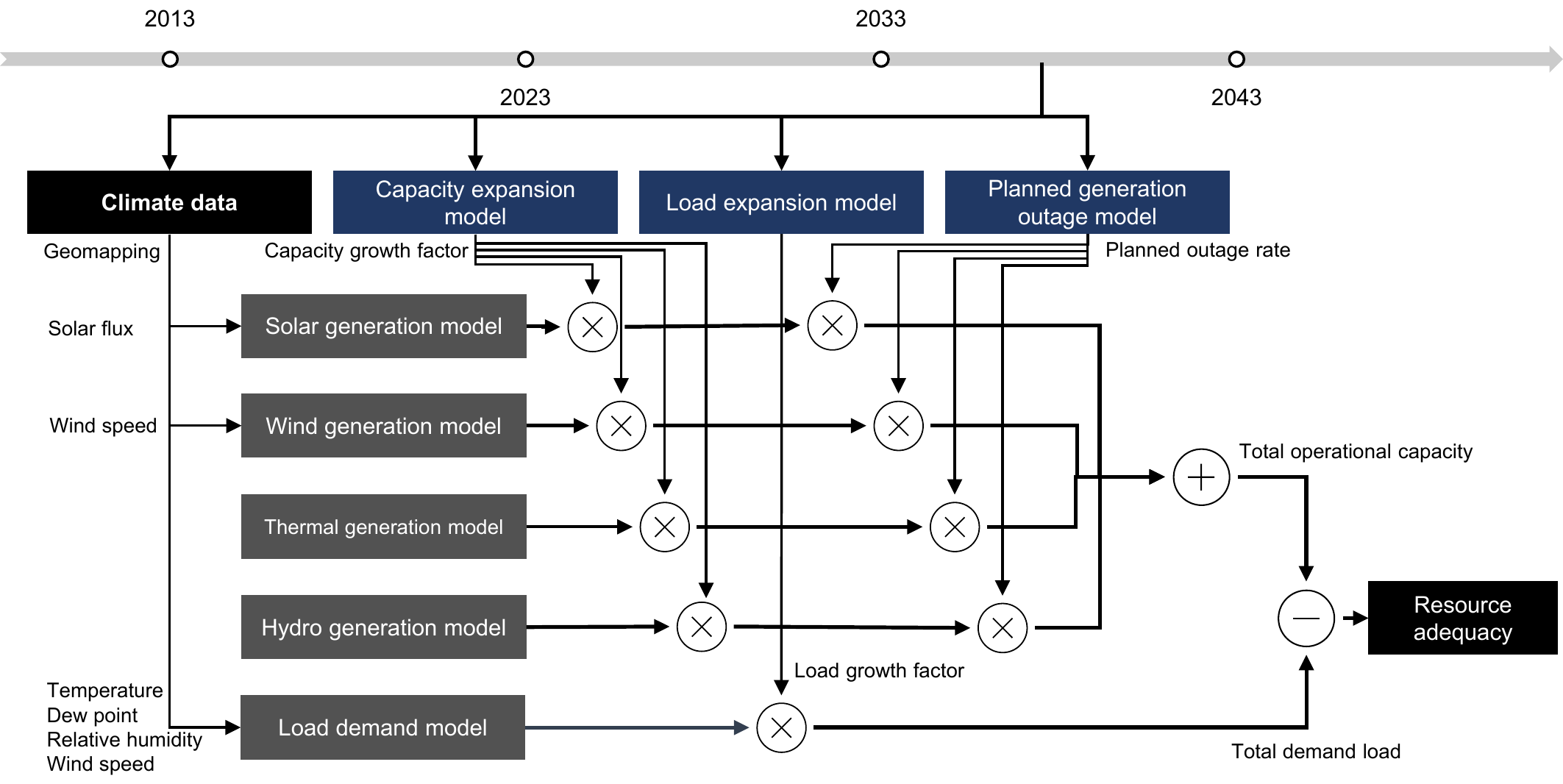}
    \caption{Schematic of resource adequacy assessment based on climate data using models of generation capacity, load, planned outage, and expansion.}
    \label{fig:diagram}
\end{figure*}

\section{Formulation of Resource Adequacy}\label{sec:problem_formulation}
In this study, we use the reserve margin, one of the resource adequacy indicators, as the reliability criterion, which equals the difference between the total concurrent load and the total operational capacity. To investigate the impact of climate simulation resolution on grid reliability, we perform reliability assessment in a bottom-up way as defined in Eq.~\ref{eq:reserve_margin}.
\begin{subequations}\label{eq:reserve_margin}
\begin{align}
&\mathbf{M}_{ymdh} = \bar{\mathbf{P}}_{ymdh}^{\text{g}}-{\mathbf{P}}_{ymdh}^{\text{l}},\\
&\mathbf{P}_{ymdh}^{\text{l}}=\sum_{i=1}^{n_\text{z}}{P}_{zymdh}^{\text{l}},\\
&\bar{\mathbf{P}}_{ymdh}^{\text{g}}=\sum_{i=1}^{n_\text{s}}\bar{P}_{iymdh}^{\text{s}}+\sum_{i=1}^{n_\text{w}}\bar{P}_{iymdh}^{\text{w}}+\sum_{i=1}^{n_\text{s}}\bar{P}_{iymdh}^{\text{t}},
\end{align}
\end{subequations}
where $M_{ymdh}$ is the reserve margin at hour $h$ on day $d$ in month $m$ of year $y$, $\bar{\mathbf{P}}_{ymdh}^{\text{g}}$ is the total operational generation capacity, $\mathbf{P}_{ymdh}^{\text{l}}$ is the total load demand. In addition, $\bar{P}_{iymdh}^{\text{l}}$ represents the zonal load demand, while $\bar{P}_{iymdh}^{\text{w}}$, $\bar{P}_{iymdh}^{\text{s}}$, and $\bar{P}_{iymdh}^{\text{t}}$ are the operational generation capacity of individual wind, solar, and traditional (thermal and hydro) plants, respectively.

In this study, we focus on evaluating the resource adequacy of the Texas interconnection between 2033 and 2043.
To achieve accurate long-term assessments of operational generation capacity and load demand, we consider not only climate factors, but also capacity and load expansion factors, as well as planned generation outages in the Texas interconnection. As a result, we require models to simulate generation capacity and load demand that can account for climate conditions, as well as models for long-term capacity expansion, load expansion, and planned generation outages.

\section{Methodology}\label{sec:Methods}
This section describes the development of a future Texas grid that encompasses precise spatial data on the location and capacity of generations and loads, along with the planning of generation outages. The constructed grid can be utilized to evaluate the system reliability using different resolutions of climate simulation data as input. The overview of the entire process is shown in Fig.~\ref{fig:diagram}.

\subsection{Data Collection and Processing}
\subsubsection{Climate Simulation Data}
Global climate simulations were conducted using the Community Earth System Model (CESM) version 1.3, with a 6-hour temporal resolution, in both high-resolution (0.25 degree or 25 km) and low-resolution (1 degree or 100 km) configurations. As discussed in~\cite{chang2020unprecedented}, high-resolution climate simulations yield significant improvement in simulating not only regional climate mean states, but also weather extreme events. To assess the impact of climate model resolution on energy system reliability, we conduct a pilot case study from 2033 to 2043 in the Texas region, utilizing atmosphere temperature, relative humidity, and dew point at 2-meter height, wind speed at 100-meter and direct and net solar radiation flux at surface obtained from the climate model output stored on a data server~\cite{CESM} to derive energy grid reliability.

\subsubsection{Data for Generation Capacity Models}
We obtain the layout and size information of individual operable generation units in the real Texas grid from the Energy Information Administration (EIA)~\cite{EIA860form}. These data contain geographical location, technology, fuel type, turbine model, nameplate power factor, and nameplate capacity over seasons.

\subsubsection{Data for Load Models}
We obtain the layout and size information of individual loads from a 2,000-bus synthetic Texas grid~\cite{birchfield2017grid} due to the lack of real high-resolution load data. This information includes longitude, latitude, and active power. In addition, we calculate the share of individual loads in their respective load zones based on their active power and total power in the load zones to obtain a more detailed understanding of the distribution of loads in the region.

Furthermore, since we will use a regression model for load modeling, it is necessary to have historical weather and load data for training and validation purposes.
To this end, we collect historical weather data at 203 Automated Surface Observing System (ASOS) sites throughout Texas from~\cite{ASOS1min}, as well as zonal load data from the Electric Reliability Council of Texas (ERCOT)~\cite{ERCOTportal}. The historical weather data contains temperature, dew point, relative humidity, and wind speed, which we average and interpolate to unify into hourly data. The zonal load data provides hourly electricity consumption at the zonal level in the real Texas grid.

\subsubsection{Data for Load Expansion Models}
As we do not utilize bottom-up econometric approaches for long-term load expansion between 2033 and 2043 in this study, we use the long-term load forecast between 2023 and 2032 collected from ERCOT~\cite{ERCOTlongtermloadforecast}.
The long-term load forecast data comprises monthly peak total load projections in the Texas interconnection from 2023 to 2032, which are generated from econometric models incorporating various factors such as the number of premises in different customer classes, weather variables, and calendar variables.

\subsubsection{Data for Planned Outage Models}
In order to examine the pattern of planned outages for various types of generation, we obtain historical data on planned generation capacity outages from ERCOT~\cite{ERCOTportal}. This data provides hourly planned capacity outages of intermittent (wind and solar) and non-intermittent (thermal and hydro) resources, respectively.

\subsection{Power Grid Modeling and Simulation}

\subsubsection{Models of Operational Generation Capacity}
We utilize physical models to estimate the operable generation capacity of each type of generation unit over the years, considering the model projected climate conditions. To obtain the climate data for a particular generation plant, we select the sample with climate data that is geographically closest in terms of Euclidean distance.

We model the operational capacity of wind farms as a function of the wind speed as defined by
\begin{equation}\label{eq:wind}
\bar{P}_{iymdh}^{\text{w}}=\left(1-\eta_{iymdh}^{\text{w}}\right)*\phi_i^{\text{w}}*\bar{P}_{iy}^{\text{w}}*C_i\left(V_{ymdh}\right),
\end{equation}
where $\eta_{iymdh}^{\text{w}}$ is the planned generation outage rate, $\phi_i^{\text{w}}$ is the nameplate power factor, $\bar{P}_{iy}$ is the nameplate capacity in year $y$, $C_i$ is the wind-speed-power curve dependent on the brand and turbine type, and $V_{iymdh}$ is the real-time local wind speed.

We model the operational capacity of solar photovoltaic (PV) farms as a function of net solar flux as defined by
\begin{subequations}\label{eq:solar}
\begin{align}
&\bar{P}_{iymdh}^{\text{s}}=\left(1-\eta_{iymdh}^{\text{s}}\right)*\phi_i^{\text{s}}*\bar{P}_{iy}^{\text{s}}*\Tilde{S}_{iymdh},\\
&\Tilde{S}_{iymdh} = {S}_{iymdh} / \max_{m,d,h}S_{iymdh},
\end{align}
\end{subequations}
where $\eta_{iymdh}^{\text{s}}$ is the planned generation outage rate, $\phi_i^{\text{s}}$ is the nameplate power factor, $\bar{P}_{iy}^{\text{s}}$ is the nameplate capacity in year $y$, $S_{iymdh}$ is the real-time local net solar flux, and $S^{\text{max}}_{iy}$ is the maximum net solar flux in year $y$.

We model the operational capacity of traditional hydro and thermal plants, including natural gas, nuclear, coal, and biomass, as a season-dependent function as defined in
\begin{equation}\label{eq:traditional}
\bar{P}_{iymdh}^{\text{t}}=\left(1-\eta_{iymdh}^{\text{t}}\right)*\phi_i^{\text{t}}*\bar{P}_{iy}^{\text{t}}*\beta_m,
\end{equation}
where $\eta_{iymdh}^{\text{t}}$ is the planned generation outage rate, $\phi_i^{\text{t}}$ is the nameplate power factor, and $\bar{P}_{iy}^{\text{t}}$ is the nameplate capacity in year $y$. Additionally, $\beta_m$ is a month-dependent factor that ranges from 0 to 1, indicating the varying efficiency in different seasons, and remains constant throughout the months within the same season.

\begin{table*}
    \centering
\begin{threeparttable}
\centering
\captionof{table}{Performance of zonal load regression models}
\begin{tabular}{ccccccccc}
    \hline
    Metric & COAST & EAST & FWEST & NORTH & NCENT & SOUTH & SCENT & WEST\\
    \hline
    R$^2$ & 0.97 & 0.97 & 0.96 & 0.97 & 0.97 & 0.97 & 0.97 & 0.97 \\
    MAE (GW) & 0.34 & 0.05 & 0.04 & 0.02 & 0.48 & 0.11 & 0.22 & 0.03 \\
    MSE (GW$^2$) & 0.21 & 3.8e-3 & 3.2e-3 & 9.1e-4 & 0.41 & 0.02 & 0.09 & 1.8e-3 \\ 
    \hline
\end{tabular}
\begin{tablenotes}
\item[$*$]Neural network configuration contains two hidden layers of 100 neurons with ReLU as activation functions.
\end{tablenotes}
\label{table:load_training}
\end{threeparttable}
\end{table*}

\begin{figure*}[hbt!]
    \centering
    \includegraphics[width =0.95\textwidth]{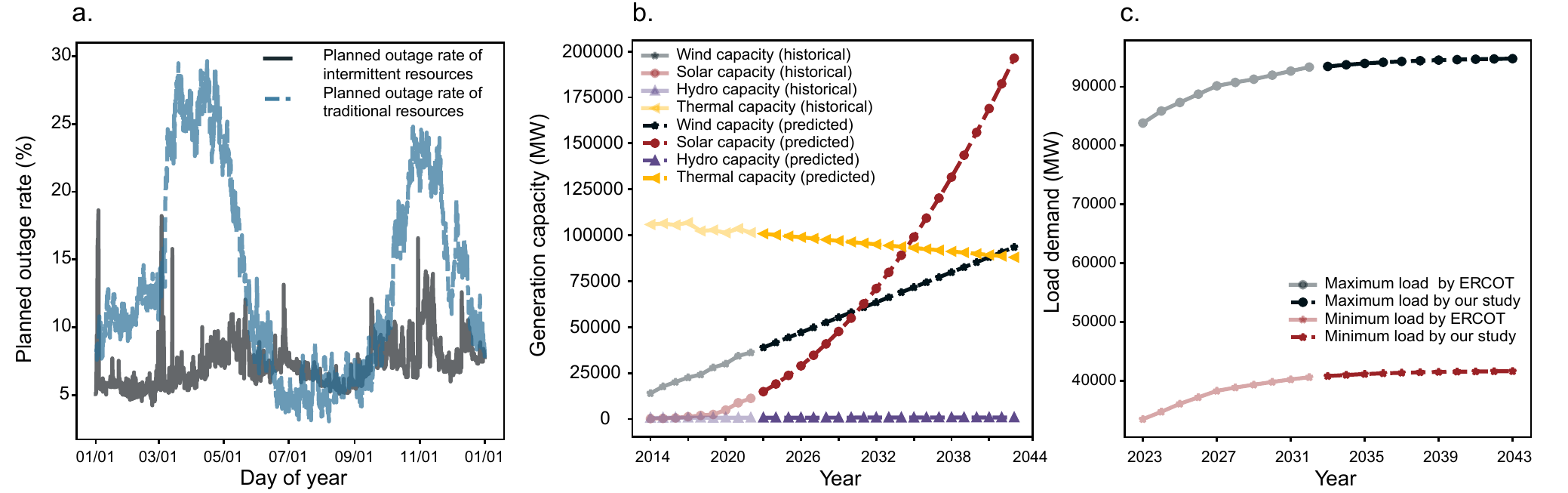}
    \caption{Planned generation outage rate, predicted generation capacity, and predicted load demand. a. Planned generation outage rates of intermittent and non-intermittent generation resources. b. Predicted generation capacity of thermal, hydro, wind, and solar generation. c. Predicted maximum and minimum loads.}
    \label{fig:expansion_outage_models}
\end{figure*}

\subsubsection{Zonal Load Model}
We utilize regression models to estimate the load demand of each load zone over the years, considering the climate conditions. Similarly, to obtain the climate data for a particular load, we select the sample with climate data that is geographically closest in terms of Euclidean distance.

We model zonal loads as functions of time, temperature, dew point, relative humidity, and wind speed as defined in
\begin{subequations}\label{eq:load}
\begin{align}
&P_{zymdh}^{\text{l}}=f_{z}\left(\Tilde{X}_{zymdh}\right)\left(P_{zy}^{\text{l},\max}-P_{zy}^{\text{l},\min}\right)+P_{zy}^{\text{l},\max},\\
&\Tilde{X}_{zymdh} = \sum_{i\in z}\alpha_{zi}{X}_{iymdh},\\
&{X}_{iymdh} = [{T}_{iymdh}, D_{iymdh}, H_{iymdh}, V_{iymdh}, m, d, h],
\end{align}
\end{subequations}
where $P_{zymdh}^{\text{l}}$ is the real-time zonal load demand, $f_{z}$ is the normalized zonal load model, $P_{zy}^{\text{l},\max}$ and $P_{zy}^{\text{l},\min}$ are the maximum and minimum load demand in zone $z$ in year $y$, $\Tilde{X}_{zymdh}$ is the average environmental condition in zone $z$, $\alpha_{zi}$ means the share of load $i$ in zone $z$ that follows $\sum_{i\in z} \alpha_i=1$, and ${T}_{iymdh}$, ${D}_{iymdh}$, ${H}_{iymdh}$, and ${V}_{iymdh}$ are local temperature, dew point, relative humidity, and wind speed at load $i$, respectively,

Specifically, we employ a neural network to implement the normalized zonal load model $f_{z}$ for each zone. In the training process, we collect the historical zonal load data as $P_{zymdh}^{\text{l}}$, calculate $\Tilde{X}_{zymdh}$ based on historical weather data, and train the model by optimizing on the objective function
\begin{equation}\label{eq:load_regression}
\min_\theta\, \sum_{y,m,d,h}\left(P_{zymdh}^{\text{l}}-f_{z}\left(\Tilde{X}_{zymdh};\theta\right)\right),
\end{equation}
where $\theta$ is the trainable parameters of the neural network.
Table~\ref{table:load_training} displays the performance of all zonal load regression models, which indicates the accuracy of load prediction.

\subsubsection{Planned Generation Unit Outage Model}
To estimate the operable capacity, we consider the planned outages of generation units. Our approach assumes that the planned outage capacity of a specific generation type is proportional to its total installed capacity. Therefore, it is essential to determine the planned outage rate, which represents the ratio of the planned outage capacity to the total installed capacity of the same generation type. We estimate the planned outage rate using the hourly planned outage capacity reported by ERCOT in 2019 and the installed capacity data from EIA in 2019.

The distribution of planned outage rates for intermittent renewable and non-intermittent traditional generation units is shown in Fig.~\ref{fig:expansion_outage_models}-a. It is observed that non-intermittent units, mainly thermal and hydro generation, have planned outages during the spring and winter of each year for maintenance purposes. On the other hand, intermittent units have relatively low rates of planned outages with a more uniform distribution.

\subsubsection{Expansion Model}
In this subsection, we will present models for capacity and load expansion that can be applied to calculate the nameplate capacity $\bar{P}_{iy}^{\text{x}}$ in Eq.~\ref{eq:wind}-\ref{eq:traditional} and the maximum $P_{zy}^{\text{l},\max}$ and minimum load demand $P_{zy}^{\text{l},\min}$ in Eq.~\ref{eq:load}.

We employ polynomial regression models to forecast the installed generation capacity for each generation type from 2023 to 2043, using the EIA's installed capacity data from 2014 to 2022. For thermal, hydro, and wind generation, we utilize first-order polynomial models (linear regression), while for solar generation, we use a second-order polynomial regression model (quadratic regression). This is primarily due to the clear superlinear increasing trend observed in the historical installed capacity of solar generation within the Texas interconnection.

Fig.~\ref{fig:expansion_outage_models}-b illustrates the trends in generation capacity for different sources. Specifically, the thermal generation capacity is expected to gradually decrease from 101,286 MW in 2022 to 94,397 MW in 2033, and eventually reach 87,983 MW in 2043. On the other hand, the hydro generation capacity is predicted to increase slowly from 723 MW in 2022 to 784 MW in 2033, and eventually reach 848 MW in 2043. The solar generation capacity is expected to experience a massive increase from 4,880 MW in 2022 to 79,795 MW in 2033, and eventually reach 196,368 MW in 2043. Finally, the wind generation capacity is expected to steadily increase from 30,059 MW in 2022 to 66,209 MW in 2033, and eventually reach 93,522 MW in 2043.

We employ regression models to estimate the trend in power demand between 2033 and 2043, using the ERCOT long-term load forecast report~\cite{ERCOTlongtermloadforecast} that predicts monthly peak loads between 2023 and 2032. We notice a trend of continuous increase with a decreasing rate of growth, and that the maximum and minimum monthly peak loads show differentiable growth rates. To account for these characteristics, we employ logistic growth regression models for regression and prediction on maximum and minimum monthly peak loads, using the equation $y=c/\left(1+a*\exp\left(-bt\right)\right)$, where $a,\,b,\,c$ are the coefficients to learn, $t$ represents the year, and $y$ denotes the maximum or minimum monthly peak load. We estimate the minimum and maximum load demand ($P_{zy}^{\text{l},\max}$ and $P_{zy}^{\text{l},\min}$) by using the predicted maximum monthly peak loads for the former, and by estimating the latter through the growth rate of the predicted minimum monthly peak loads and historical minimum load data from 2019.

Fig.~\ref{fig:expansion_outage_models}-c depicts the trend in maximum and minimum loads. It is worth emphasizing that the minimum and maximum loads for future years are used for scaling purposes and should be interpreted as the anticipated minimum and maximum load under the same weather conditions as in the past, rather than as the precise minimum and maximum loads based on future climate projections. It is observed that the growth rate of the maximum load is 13\% between 2023 and 2043, whereas that of the minimum load is 24\%. Specifically, the maximum peak load will increase from 82,308 MW in 2023 to 91,755 MW in 2033, and eventually reach 93,054 MW in 2043, while the minimum load will increase from 33,517 MW in 2023 to 40,838 MW in 2033, and eventually reach 41,668 MW in 2043.

\section{Results}\label{sec:results}

\subsection{Resource Adequacy Assessment}\label{subsec:reliability}
We utilize the proposed models for load demand and operational generation capacity to estimate the resource adequacy from 2033 to 2043, using data from single high and low-resolution climate simulations, respectively. For the sake of clarity, we will refer to the reliability assessment based on the high-resolution (low-resolution) climate simulation data as the high (low) resolution reliability assessment in the rest of this section.

\begin{figure}[btp!]
    \centering
    \includegraphics[width = 1\columnwidth]{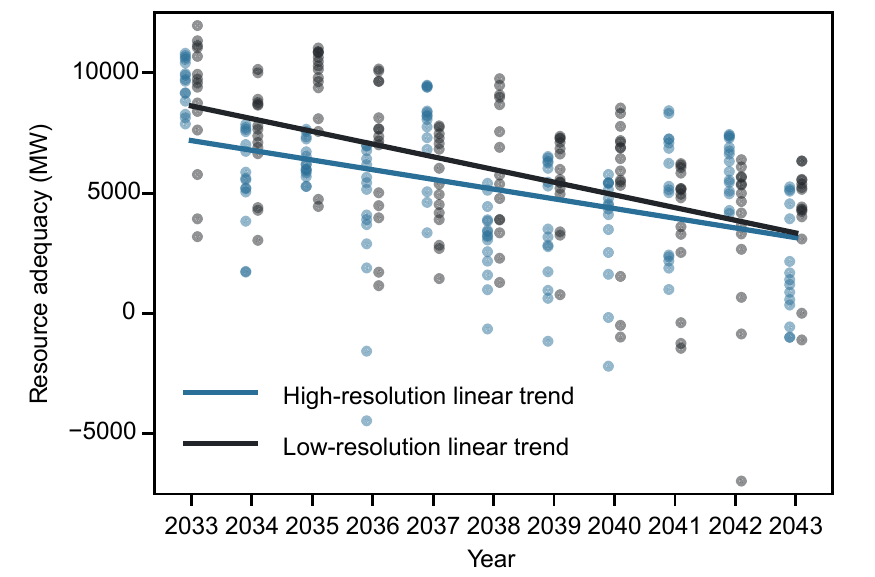}
    \caption{The linear trends of the average adequacy for the top 1\% of events with the lowest adequacy, as determined by both high and low-resolution assessments.}
    \label{fig:adequacy_trend}
\end{figure}

Since analyzing climate-induced extreme events is a crucial objective of long-term reliability assessment, we initially prioritize the top 1\% of events with the lowest adequacy in each year between 2033 and 2043.
Fig.~\ref{fig:adequacy_trend} illustrates the linear trend of their average resource adequacy, with both high and low-resolution reliability assessments indicating a clear decreasing trend over the years. Additionally, the high-resolution assessment consistently displays lower adequacy than the low-resolution assessment.

\begin{figure}[bt!]
    \centering
    \includegraphics[width = 1\columnwidth]{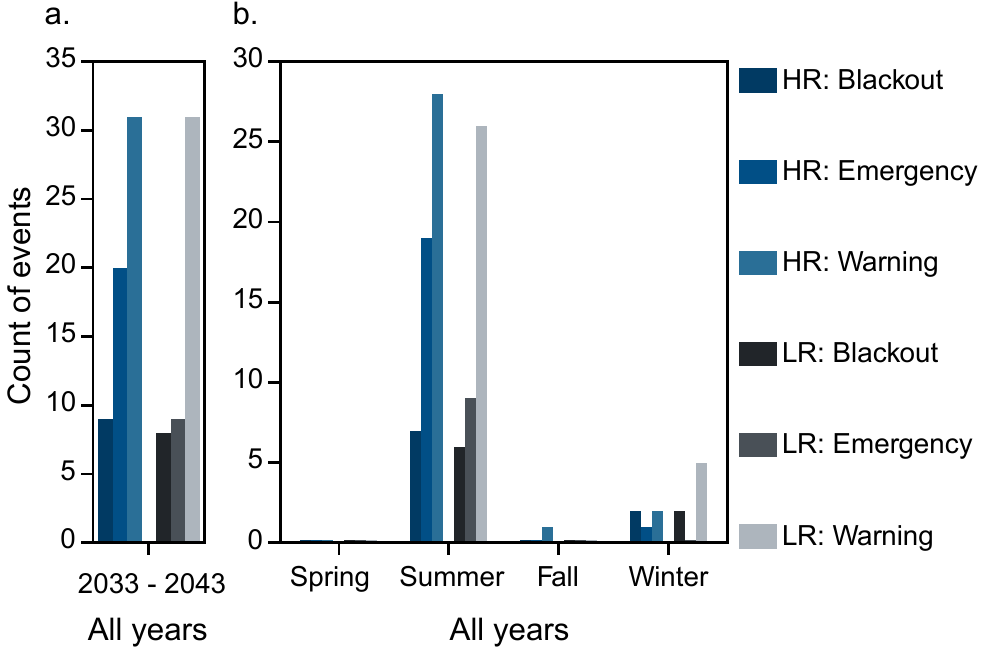}
    \caption{Counts of blackout, emergency, and warning events between 2033 and 2043 based on high (HR) and low (LR) resolution reliability assessment. a. Total count of each type of events from 2033 to 2043. b. Total count of each type of events in each season from 2033 to 2043.}
    \label{fig:adequacy_barplot_year_season}
\end{figure}

\subsection{Statistics of Extreme Events}
To achieve a better understanding of potential climate-induced extreme events, we classify low-adequacy extreme events into three types based on the threshold of the Energy Emergency Alert (EEA) state, which is triggered when the reserve in the Texas interconnection falls below 2,300 MW. These event types are: (1) blackout events, which happen when the resource adequacy falls below 0, (2) emergency events, which occur when the resource adequacy ranges from 0 to 2,300 MW, and (3) warning events, which take place when the resource adequacy ranges from 2,300 to 4,600 MW.


It is evident from Fig.~\ref{fig:adequacy_barplot_year_season}-a that high-resolution reliability assessment can detect more emergency events between 2033 to 2043, but there is no significant difference in the number of blackout and warning events detected. To be precise, the high-resolution assessment identifies a total of 60 extreme events, while the low-resolution assessment identifies 48.
Notably, the seasonal distribution of the counts of events shown in Fig.\ref{fig:adequacy_barplot_year_season}-b shows that the grid experiences more frequent events during the summer months, which aligns with empirical observations. Furthermore, the high-resolution assessment detects a greater number of extreme events in all three categories. Specifically, the high-resolution assessment shows a significantly higher number of emergency events during the summer months. Please see more detailed statistics of extreme events in Appendix~\ref{appendix:statistics of extreme events}.


\begin{figure}[bt!]
    \centering
    \includegraphics[width =1\columnwidth]{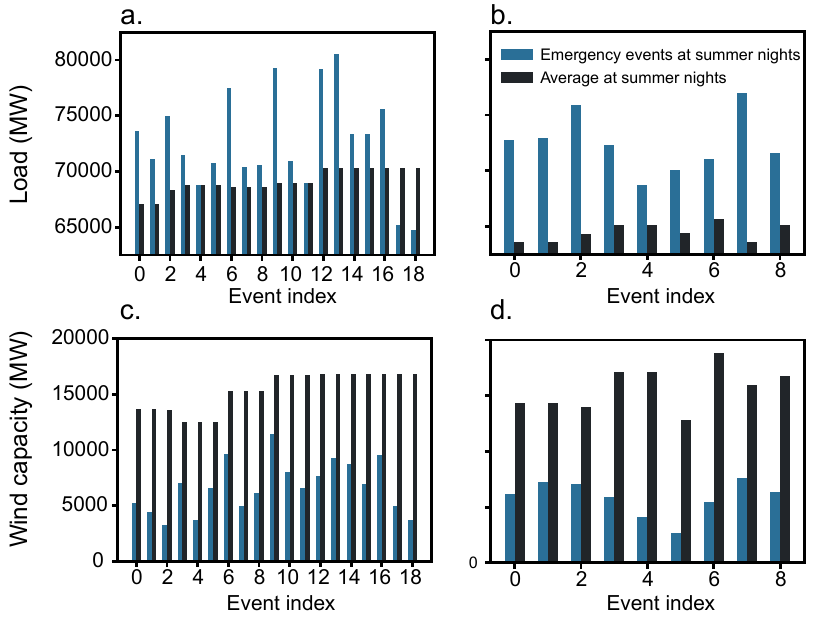}
    \caption{Comparison of load power and wind capacity between summer emergency events and corresponding summer night averages in simulated data. a, b show show the comparison of load power between emergency events and summer night average for high and low-resolution assessments, respectively. c, d show the comparison of wind capacity between emergency events and summer night average for high and low-resolution assessments, respectively.}
    \label{fig:summer_emergency}
\end{figure}

\begin{figure}[bt!]
    \centering
    \includegraphics[width =1\columnwidth]{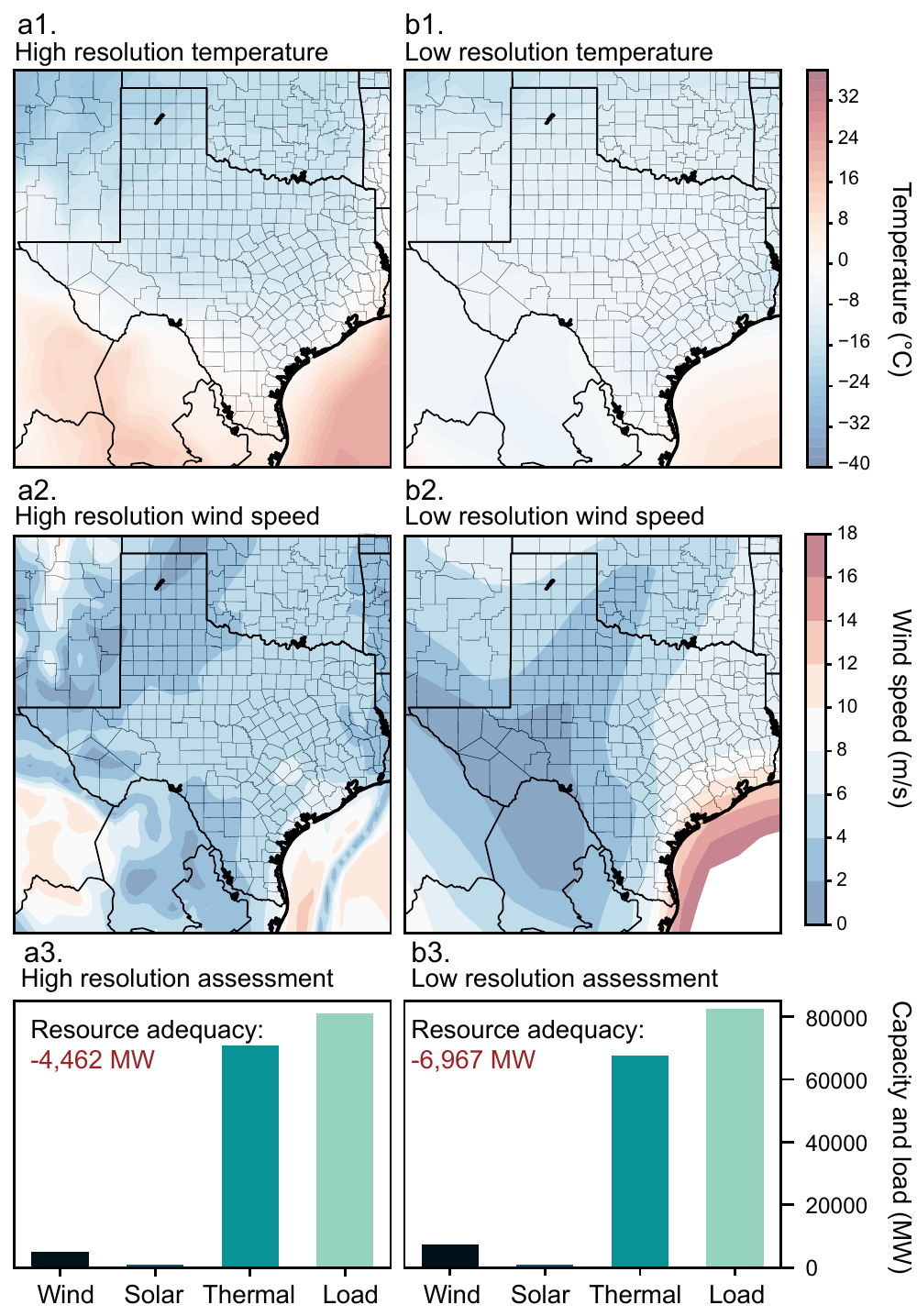}
    \caption{Simulated climate conditions during two worst blackout events detected by high and low-resolution reliability assessments. a1, a2, a3 show the high-resolution simulated temperature, wind speed, and the corresponding estimated generation mix and adequacy during 0 to 6 a.m. on February 5$^\text{th}$, 2036. b1, b2, b3 show the low-resolution simulated temperature, wind speed, and the corresponding estimated generation mix and adequacy during 0 to 6 a.m. on January 10$^\text{th}$, 2042.}
    \label{fig:temperature_extreme_event}
\end{figure}

As there are notable disparities between the summer emergency events detected by high and low-resolution assessments, we investigate the causes of these events. Surprisingly, all of these events occur during the night period from 6 p.m. to 12 a.m. While the low-resolution assessment identifies emergency events caused by consistently high load and low wind capacity compared to the average during summer nights (Fig.~\ref{fig:summer_emergency}-b and d), the high-resolution assessment reveals more emergency events with two types of causes (Fig.~\ref{fig:summer_emergency}-a and c), including (1) high load and low wind capacity and (2) average or below-average load with extremely low wind capacity. This suggests that high-resolution climate simulation provides a wider range of weather conditions that can lead to extreme events with low adequacy.

To summarize, while low-resolution climate simulation can aid in a first-order long-term system reliability assessment of the Texas interconnection between 2033 and 2043, high-resolution climate simulation is more effective in providing informative assessments, especially for emergency events that occur during the night in summer with low wind speeds.

\subsection{Blackout Events in Winter}\label{subsec:the worst blackout events}
Despite blackout events being more frequent during the summer, their severity during winter is even greater than that of summer, as measured by the average resource adequacy. In this context, we present the worst blackout events identified by high and low-resolution reliability assessments for further discussion.
Specifically, the worst blackout event detected by the high-resolution assessment occurs during the early morning hours of February 5$^\text{th}$, 2036, with a power shortage of 4,462 MW, while the worst event detected by the low-resolution assessment occurs during the early morning hours of January 10$^\text{th}$, 2042, with a power shortage of 6,967 MW.
It is important to note that the timing information presented in the data is for illustrative purposes only, as climate simulations cannot provide precise timing information for any single extreme event. Moreover, the relative severity between the two blackout events should not be taken as a conclusion as well, as it is based on observations from single simulations. Please see the detailed explanation in Appendix~\ref{appendix:compare extreme events}.

Fig.~\ref{fig:temperature_extreme_event} shows comparison of high and low-resolution simulated climate conditions during the two identified worst blackout events. 
The common cause of both events is the compound effects of the high load demand due to extremely low temperatures (Figs.~\ref{fig:temperature_extreme_event}-a1 and b1), low wind generation power resulting from low wind speeds (Figs.~\ref{fig:temperature_extreme_event}-a2 and b2), and lack of solar generation power during early morning hours, as shown in Figs.~\ref{fig:temperature_extreme_event}-a3 and b3.


Despite determining the severity of the worst blackout event through a single round of simulation may be not meaningful because of the small sample size (Appendix~\ref{appendix:compare extreme events}), there may still be important policy implications that can be derived from both scenarios. First, the grid will continue to face challenges in meeting demand during low-temperature extremes, even with over 200,000 MW of installed renewable generation. This highlights the importance of developing complementary technologies to address the intermittency of renewables, such as large-scale demand response programs and long-term energy storage. Second, our model does not account for the possibility of generation facility malfunction during extreme climate events, such as the extremely low temperatures experienced in Texas in February 2021. This can greatly exacerbate grid reliability issues. Therefore, implementing weatherization measures for all generation units is crucial to enhance grid reliability against extreme weather, as such events will not be isolated incidents in the future.

\subsection{Discussion}
In this study, the extendable framework that can incorporate granular climate data allows higher resolution modules of operation and long-term planning. For instance, future studies can customize high-resolution modeling of emerging technologies for better assessment, such as granular layout and sizing information of massive roof-top solar PV panels, long-term energy storage, large flexible loads, and fast growing electric vehicles (EVs). By leveraging state-of-the-art climate models and accounting for the inherent uncertainties, researchers and policymakers can more effectively assess the potential impacts of climate change on energy grid reliability and develop suitable adaptation and mitigation strategies to ensure a resilient and sustainable energy future.
Despite the high spatial resolution of climate data and the extendability of the assessment framework, it is crucial to acknowledge that this study has several significant areas that require improvement in future research.

Regarding climate simulation output, it would be advantageous to have hourly data with a large ensemble size to capture more statistically reliable information. This pilot study incorporating high-resolution climate modeling with the power system model could be further optimized by increasing the output frequency from every 6 hours to every hour. In solar-dominated grids, such as the California interconnection, resource scarcity can be a challenge during sunset hours, while in wind-dominated grids, wind speeds can vary significantly within an hour. By obtaining hourly climate simulation outputs, we can gain a better understanding of the variability of renewable resources and accurately assess their impact on the power grid, aiding in informed decisions regarding optimal siting and sizing. Furthermore, a single realization of climate simulation as we used in this pilot study cannot adequately assess the statistical robustness and uncertainties of future climate, and a large ensemble of high-resolution climate simulations is desirable for future studies. With large ensemble simulations, we can obtain more robust statistics about extreme events, allowing us to examine in more details the impact of different model resolutions.

Regarding power system modeling, this study has several simplifications that need improvement in future research. First, for long-term load modeling, we have not taken into account emerging technologies such as rooftop solar PV panels and EVs, which can significantly alter the future energy consumption pattern. Second, we have not used an econometric model to predict the development of demographics and sectors that can change the distribution of load demand across Texas. Third, we have not used geophysical models to assess local renewable potential to guide the deployment and retirement of traditional and renewable generation plants. Additionally, we have not used binary outage states or derates for each individual power plant, which needs further refinement. Finally, we have not incorporated unit commitment and economic dispatch to represent the impact of temporal correlated decision-making and the limitation of system topology, such as congestion.

\section{Conclusion}\label{sec:conclusion}
The presented framework in this pilot study provides a preliminary long-term reliability assessment for the Texas interconnection from 2033 to 2043, utilizing both high and low-resolution climate simulation data. Our findings reveal that high-resolution climate data can more accurately identify emergency events, particularly those that occur during summer nights. Furthermore, the analysis shows that both high and low-resolution assessments detect the worst blackout event in winter. Future studies with large ensemble simulations are needed to validate the statistical robustness of the conclusions. There are several areas for future research, such as the need for hourly climate simulation outputs, integration of emerging technologies and demographics, and the incorporation of renewable potential and unit commitment in power system modeling.

\appendices
\section{Statistics of extreme events over years}\label{appendix:statistics of extreme events}
\begin{figure}[h]
    \centering
    \includegraphics[width =1\columnwidth]{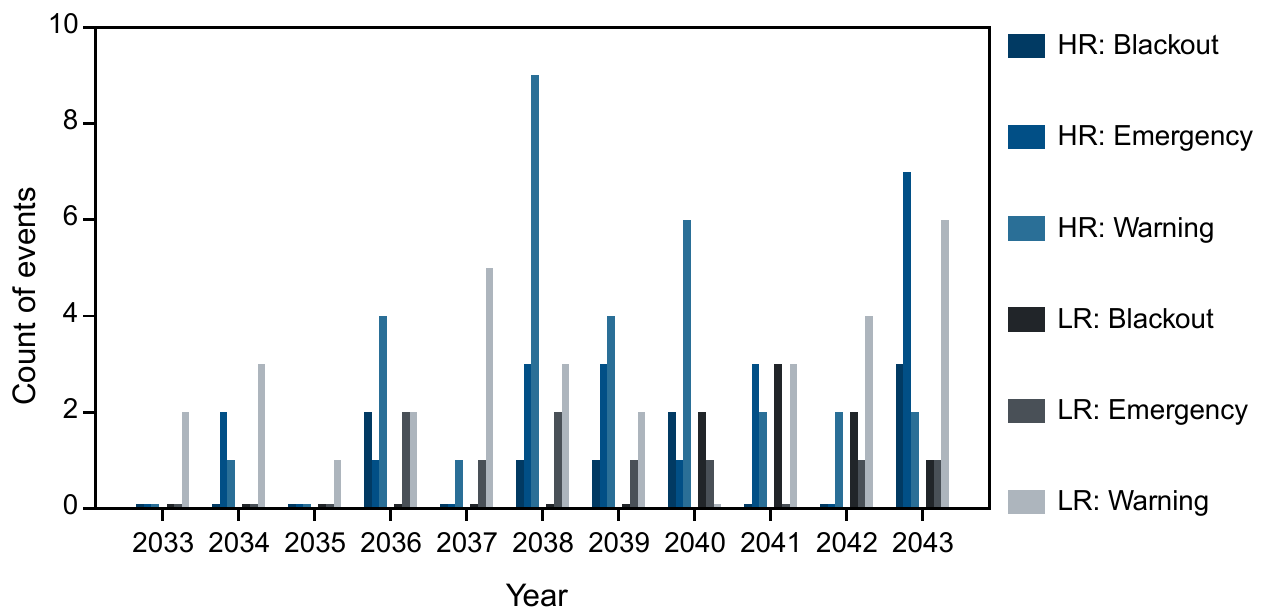}
    \caption{Counts of blackout, emergency, and warning events for each year between 2033 and 2043 based on high (HR) and low (LR) resolution reliability assessment.}
    \label{fig:appendix adequacy_barplot_annual}
\end{figure}
\begin{figure*}[tb!h]
    \centering
    \includegraphics[width =1\textwidth]{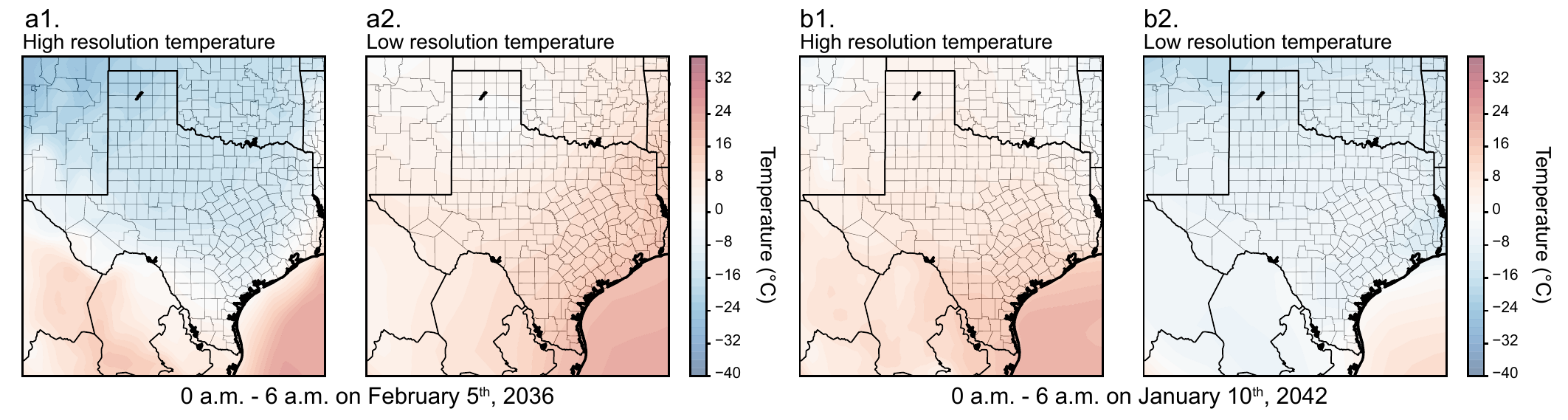}
    \caption{Comparison of simulated temperatures during the worst blackout event detected by high-resolution reliability assessment, which occurs during 0 to 6 a.m. on February 5$^\text{th}$, 2036, as well as the worst blackout event detected by low-resolution reliability assessment, which occurs during 0 to 6 a.m. on January 10$^\text{th}$, 2042. a1, a2 show the simulated contrasting temperatures in the worst event detected by high-resolution reliability assessment during the first blackout event. b1, b2 show the simulated contrasting temperatures in the worst event detected by low-resolution reliability assessment during the second blackout event.}
    \label{fig:appendix - temperature_extreme_event}
\end{figure*}
Fig.~\ref{fig:appendix adequacy_barplot_annual} shows the counts of different types of events from 2033 to 2043 by high and low-resolution reliability assessment. Interestingly, there is no consistent overestimation or underestimation in the relative relationship between the yearly counts of events determined by the low-resolution climate data and those determined by the high-resolution climate data. This is primarily due to the fact that the difference between high and low-resolution climate modeling cannot be solely attributed to differences in granularity that can be approximated by interpolation. Instead, there may be qualitative differences between the two approaches. In other words, the low-resolution climate simulation may show extreme weather during certain periods that are not present in the high-resolution simulation, and vice versa.

\section{Differences between high and low-resolution climate simulations}\label{appendix:compare extreme events}
It is important to emphasize that the difference between low and high-resolution climate simulations at the same period is not merely a matter of increased granularity that can be approximated by interpolation, but rather a fundamental qualitative difference. This is demonstrated in Fig.~\ref{fig:appendix - temperature_extreme_event}, which compares the simulated temperatures during the worst blackout events in winter using both high and low-resolution assessments. The climate simulation process essentially involves simulating stochastic partial differential equations with slightly perturbed initial states. Therefore, different rounds of simulations, even with the same resolution configuration, may result in completely different outcomes in each period. Hence, analyzing a single event by a single round of simulation is not appropriate. To obtain useful and reliable information, it is essential to analyze the statistics of certain variables over a sufficiently long time horizon. Additionally, it is critical to increase the number of simulations in the ensemble to obtain statistically robust conclusions.

\section*{Acknowledgment}
This research is partly supported by NSF grant AGS-2231237. Portions of this research were conducted with the advanced computing resources provided by the Texas Advanced Computing Center at University of Texas, Austin and Texas A\&M High Performance Research Computing.

\ifCLASSOPTIONcaptionsoff
  \newpage
\fi

\bibliographystyle{IEEEtran}

\end{document}